\documentclass{ws-ijmpb}
\usepackage{amsfonts,amssymb,amsmath}
\usepackage{bm,euscript,mathrsfs}
\usepackage[utf8]{inputenc}
\usepackage[T1]{fontenc}
\usepackage{color,soul}
\usepackage{textcomp}
\usepackage[tight,sf]{subfigure}
\usepackage{hyperref}
\usepackage{graphicx}
\usepackage{color}
\usepackage{psfrag}

\begin{document}

\markboth{J. A. Santiago, G. Chac\'on-Acosta and O. Gonz\'alez-Gaxiola}
{Elastic  curves and surfaces under long-range\\
forces: A geometric approach}

%
\catchline{}{}{}{}{}
%

\title{ELASTIC  CURVES AND SURFACES UNDER LONG-RANGE\\
 FORCES: A GEOMETRIC APPROACH  }

\author{J. A. SANTIAGO${}^*$, G. CHAC\'ON-ACOSTA${}^\dag$ AND O. GONZ\'ALEZ-GAXIOLA$^{\ddag}$}

\address{Departamento de Matem\'aticas Aplicadas y Sistemas,\\ Universidad Aut\'onoma
Metropolitana-Cuajimalpa,\\ Artificios 40,
M\'exico D. F., 01120,
M\'EXICO\\
${}^*$jsantiago@correo.cua.uam.mx \\
${}^\dag$gchacon@correo.cua.uam.mx\\
${}^{\ddag}$ogonzalez@correo.cua.uam.mx}

\maketitle

\begin{history}
\received{Day Month Year}
\revised{Day Month Year}
\end{history}

\begin{abstract}
Using classical differential geometry, the problem of elastic curves  and surfaces in the presence of long-range interactions $\Phi$, is posed.
Starting from a variational principle, the balance of elastic forces and the corresponding projections  ${\bf n}_i\cdot \nabla\Phi$, are found.
In the case of elastic surfaces, a force coupling the mean curvature with the external potential, $K\Phi$,  appears;  it is also  present in the shape equation along the normal principal
in the case of curves. The potential $\Phi$ contributes to the effective tension of curves and surfaces and also to the orbital torque.
The confinement of a curve on a surface is also addressed, in such a case,  the potential  contributes to the normal force through the terms $-\kappa\Phi -{\bf n}\cdot \nabla\Phi$.
In general, the equation of motion becomes integro-differential  that must be numerically solved.
\end{abstract}

\keywords{Elastic curves and surfaces; charged polymers.}

\section{Introduction}

The elastic energy of a polymer chain, can be modeled  as a geometric functional, invariant under reparametrizations,   ${\cal H}=\int ds\, f (\kappa, \tau)$, $\kappa$
and $\tau$ being the Frenet-Serret curvatures  and $s$ the arclength of the curve accounting the polymer.
Working within the geometric framework of Frenet-Serret, an infinitesimal deformation of the curve can be decomposed in its perpendicular and tangential parts;  the invariance under reparametrizations has the effect that  just  orthogonal projections  play a role in the equilibrium shape equations  while a tangential deformation yields only a boundary term.
Is possible to integrate the Euler-Lagrange equations for the curvature in the case of plane curves \cite{langer}.
If the energy only depends  on the curvature, the model is integrable, and then the torsion is a function only of curvature\cite{arreaga}.
Furthermore, Noether conserved charges of the model, associated with invariances of motions in space, can be obtained using this geometric formalism.

In certain processes in cell biology and in certain technological applications, the behavior of polymers under the influence of external fields is of great interest.
The problem of polymers chains constrained to confined geometries was already considered \cite{pablo1}. However, there was not taken into account the role of a potential acting on the polymer. For example, in the case of DNA packaging, not only the entropic and elastic forces act on the polymer, but also the electrostatic force. Indeed, as the DNA is immersed in a solution with several other ions, the effective electrostatic potential can be modified as the Debye-H\"{u}ckel potential \cite{DH} $\Phi(r)\propto r^{-1}e^{-k r}$, that results from a linearized mean-field procedure. In general all polyelectrolytes, which are charged polymer chains, that are in solution have similar behavior. When the polymer chains are negatively charged, the charges produce their mutual repulsion and the polymer can not be wrapped, so the chain is stretched \cite{Poly}. They also have certain behavior when the chain is close to a charged surface, this behavior has been used to model some biological interactions such as in nucleosome structures of DNA~\cite{poly2}. The properties of the polyelectrolytes are well known, see e.g.
refs.~\refcite{poly1}--\refcite{rubin}, however, the geometric properties under external long-range interactions are not yet studied in detail.
In the present work we consider besides the elastic forces, the long-range interactions on curves and surfaces.

In section (\ref{Curves}), starting from a variational principle, we obtain the Euler-Lagrange equations for elastic curves in presence of long range forces.
As a result of translation and rotational invariance of the energy, the corresponding Noether charges are found. The example of elastic curves dependent on the curvature $\kappa$,
and the particular case of planar curves are shown, for these cases a first integral of the shape equation can be found.
In section (\ref{Surfaces}), the analogous analysis for surfaces with long range forces is presented. The shape equation from a variational principle and the corresponding forces and torque are found. Explicit equations in the case of fluid membranes are presented.
In section (\ref{Auxiliary}), we use the alternative method introducing auxiliary variables to regain the Euler-Lagrange equations of section (\ref{Curves}). The equations of elastic curves in presence of long range forces, constrained to surfaces
are presented.  Moreover, the normal force that constraint the curve is found in terms of the geometric information.
We conclude with some final remarks in section (\ref{Remarks}).

 \section{Curves}\label{Curves}

\subsection{Shape equations}
\label{curves}
In the presence of long range interactions, an elastic curve can be modeled by the following energy functional
\begin{equation}
{\cal H}=\int ds f(\kappa, \tau) + {1\over 2}\iint \, ds\, ds'\,  h(r).\label{modt}
\end{equation}
The first term  ${\cal H}_a= \int ds\, f(\kappa, \tau)$, constitutes the elastic energy of a curve
in ${\bf R}^3$, ${\bf x}={\bf X}(s)$,  parametrized by arclength $s$ \cite{LandauLifshitz}. The  curvature $\kappa$ and torsion $\tau$ of the curve
are defined through the  Frenet-Serret equations:
\begin{eqnarray}\label{sf}
\dot{\bf t}&=&\kappa\mathbf{n}_1,\nonumber\\
\dot{\bf n}_1&=&-\kappa\mathbf{t}+\tau\mathbf{n}_2,\\
\dot{\bf n}_2&=&-\tau\mathbf{n}_1,\nonumber
\end{eqnarray}
here the dot indicates the derivative respect to the arclength $s$.  The unit tangent vector ${\bf t}=\dot {\bf X}$, together the unit normal vectors ${\bf n}_i$, define a local basis  $\{ {\bf t}, {\bf n}_1, {\bf n}_2\} $, adapted to the curve \cite{docarmo}. Among others, some relevant cases are the energy $f(\kappa)=\kappa^2$ associated with the bending of a curve, the Euler {\it elastica} \cite{singer}, the Sadowski functional $f(\kappa)=(\kappa^2+\tau^2)^2/\kappa^2$ related with bending of elastic developable surfaces~\cite{maha}, etc.

The second term in the model (\ref{modt}) is
${\cal H}_b=\frac{1}{2}\iint\, ds\, ds' h(r)$ with $r=|{\bf X}(s)-{\bf X}(s')|$, represents the energy of the curve in the presence of long range forces,
electrostatic interaction $h(r)=1/r$ as well as the screened Coulomb interaction (Yukawa) $h(r)=e^{-\alpha r}/r$ and the Lennard-Jones potential
$h(r)=\left( \frac{\sigma}{r} \right)^{12}-\left( \frac{\sigma}{r}\right)^6$ are some relevant possibilities to explore.
The invariance under euclidean motions as well as reparametrizations of the model can be exploited to obtain several relevant properties.

In order to obtain the shape equations we have to consider an infinitesimal deformation of the energy $\delta\cal H$, as a result of deformations in the embedding functions,
 ${\bf X}\to {\bf X}+\delta{\bf X}$. For instance,
given that the arclength  is defined by $ds^2= d{\bf X}\cdot d{\bf X}$, its deformation can be written as $\delta ds=d(\delta{\bf X})\cdot {\bf t}$.
Thus, and following reference~\refcite{hamiltonian}, we can see that an arbitrary infinitesimal deformation can be written as
\begin{equation}
\delta {\cal H}_a =\int ds \hspace{2pt}\mathcal{E}_i\,\delta{\bf X}\cdot {\bf n}_i+\int ds \hspace{2pt}\dot Q,\\\label{def1}
\end{equation}
where $\mathcal{E}_i$ ($i=1,2$), are the Euler-Lagrange operators and  $Q$ the corresponding Noether charge.

Let us now consider,  the deformation of the second term in equation (\ref{modt}).  Since it is symmetric under the change of $s$ and $s'$, we have
that $\delta {\cal H}_b=\int \delta (ds) \int ds'  h(r)+ \int ds\, \delta{\bf X}\cdot\nabla \int  ds' h(r)$,  where $\nabla$ denotes the $3D$ gradient operator.
After integration by parts and introducing
the potential $\Phi:=\int ds' \, h(r)$ we can write
\begin{equation}
\delta{\cal H}_b=\int   d \left( \delta {\bf X}\cdot {\bf t}\,\Phi \right)  -\int ds\, \kappa\, \delta{\bf X}\cdot {\bf n_1}\,\Phi-\int ds\, \delta{\bf X}\cdot {\bf t}\, \dot \Phi
+\int ds\, \delta{\bf X}\cdot \nabla \Phi .\label{def2}
\end{equation}
Notice that the last term of equation (\ref{def2}) can be decomposed in its normal and tangential projections as $\delta{\bf X}\cdot \nabla \Phi =\delta{\bf X}\cdot{\bf t}\, \dot\Phi+
(\delta{\bf X}\cdot {\bf n}_i ) {\bf n}_i\cdot\nabla\Phi  $. The first term cancels the corresponding tangential projection in equation (\ref{def2}). As expected by invariance of
$\cal H$ under reparametrizations, tangential deformations do not play a role in the shape equations, these can be obtained from $\delta {\cal H}= \delta{\cal  H}_a+\delta{\cal H}_b=0$:
\begin{eqnarray}
E_1&:=&{\cal E}_1-\kappa\,\Phi+{\bf n}_1\cdot \nabla\Phi=0,\nonumber\\\label{euler}
E_2&:=&{\cal E}_2+{\bf n}_2\cdot\nabla\Phi=0.
\end{eqnarray}
The normal projections of the external force $\nabla\Phi$,  have contribution to these equations as it should be.
It is interesting to note that the potential  $\Phi$, contributes directly to the shape equation along the principal normal direction ${\bf n}_1$. This might be interpreted
as a minimal coupling between the curve and the external field.

\subsection{Translational invariance}

By adding equations (\ref{def1}) and (\ref{def2}), the total deformation of the energy $\cal H$  can  be cast as
\begin{equation}
\delta {\cal H}=\int ds \, E_i \,\delta{\bf X}\cdot {\bf n}_i +\int ds\, \dot {\cal Q}\label{defor}
\end{equation}
where now, ${\cal Q}=Q+\delta{\bf X}\cdot {\bf t}\, \Phi$, is the corresponding  Noether charge. According to equation (\ref{defor}),
under a  translation $\delta {\bf X}={\bf a}$,  the energy is deformed as
$
\delta {\cal H}={\bf a}\cdot \int ds \left( E_i \,{\bf n}_i -\dot {\bf  G} \right),
$
where
$
{\bf G}={\bf F}-{\bf t}\,\Phi,
$
being the relation with the elastic forces through  $Q= -{\bf a}\cdot {\bf F}$. In equilibrium ${\bf G}$  is a conserved vector field along the curve.
Expressing in the local basis, ${\bf G}=G_{\parallel}{\bf t}+G_i{\bf n}_i$, the balance equation $E_i \,{\bf n}_i =\dot {\bf  G}$ yields
\begin{eqnarray}
\dot G_{\parallel}-\kappa G_1&=&0,\nonumber\\
\dot G_1+\kappa G_{\parallel}-\tau G_2&=&E_1,\\
\dot G_2+\tau G_1&=&E_2,\nonumber
\end{eqnarray}
the first of them, or Bianchi identity, implies that the arclength of the curve is preserved under the total force $\bf G$. In terms of the elastic and external forces  we have
\begin{eqnarray}
\dot F_{\parallel} -\kappa F_1&=& {\bf t}\cdot\nabla \Phi,   \nonumber\\
\dot F_1+\kappa F_{\parallel}-\tau F_2&=& {\cal E}_1+{\bf n}_1\cdot \nabla\Phi,\label{ge}\\
\dot F_2+\tau\dot F_1&=&{\cal E}_2+{\bf n}_2\cdot\nabla\Phi. \nonumber
\end{eqnarray}
The first equation tell us that arclength is not preserved under the elastic force ${\bf F}$, but it is balanced with the tangential component of the external
force $\nabla\Phi$. In the same way, this force contributes to the equilibrium in the normal directions as we can see from the last two equations in (\ref{ge}).

\subsection{Rotational invariance}

The invariance under rotations  can be explored in a similar way as in ref. \refcite{hamiltonian}. Considering an infinitesimal rotation
$\delta {\bf X}=\boldsymbol{\Omega} \times {\bf X}$ in equation (\ref{defor}), we obtain
\begin{equation}
\delta {\cal H}=\boldsymbol{\Omega }\cdot \int ds\,\left( E_i\, {\bf X}\times {\bf n}_i -\dot{\bf M}\right),
\end{equation}
here,  the relation ${\cal Q}=-\boldsymbol{\Omega}\cdot {\bf M}$, was used. The equation $E_i\, {\bf X}\times {\bf n}_i =\dot{\bf M}$, follows from rotational
invariance. ${\bf M}$ is conserved along curves in equilibrium and it is identified with the torque respect to the origin.
The long range potential contributes
to the orbital torque as
\begin{equation}
{\bf M}={\bf X}\times\left(  {\bf F} - {\bf t}\,  \Phi\right) +{\bf T},
\end{equation}
where  the  local torque $\bf T$ satisfies
that $\dot{\bf T}={\bf G}\times {\bf t}$. Moreover, $\dot{\bf T}={\bf F}\times {\bf t}$ since  the contribution of the external force is in the direction of the tangent vector $\bf t$.
Therefore, decomposing $\bf T$ in the local basis, it satisfies
\begin{eqnarray}
\dot T_{\parallel}-\kappa T_1&=&0,\nonumber\\
\dot T_1 -\tau T_2+\kappa T_\parallel &=&F_2,\\
\dot T_2+\tau T_1 &=&-F_1,\nonumber
\end{eqnarray}
the external long range potential $\Phi$, does not contribute.

\subsection{Elastic curves}
Let us focus on the case where the elastic density energy is only function of the curvature $f=f(\kappa)$. The two corresponding  Euler-Lagrange operators and the
Noether charge are given by
\begin{align}
E_1 &= \ddot{f_{\kappa}}+(\kappa^2-\tau^2)f_{\kappa}-\kappa( f+\Phi)+{\bf n}_1\cdot \nabla\Phi,\label{10}\\
E_2 &=2\tau\dot{f_\kappa}+\dot\tau f_\kappa+{\bf n}_2\cdot\nabla\Phi,\label{11}\\
{\cal Q} &= (f+\Phi) \Psi_\|+f_\kappa\dot{\Psi_1}-\dot{f_\kappa}\Psi_1-2\tau f_\kappa\Psi_2,
\end{align}
where $f_\kappa=\partial f/\partial \kappa$ and the projections of the deformation, $\delta{\bf X}=\Psi_{\parallel}{\bf t}+\Psi_i{\bf n}_i$, were used.
The conserved force $\bf G$, is given by
\begin{equation}
{\bf G}=(f_\kappa\kappa -f-\Phi){\bf t}+\dot f_\kappa {\bf n}_1+\tau f_\kappa {\bf n}_2,
\end{equation}
thus, the external potential $\Phi$ contributes  to the tension in the curve. There is no contribution of the external force  to the local torque ${\bf T}$
and then ${\bf T}=-f_\kappa {\bf n}_2$ \cite{hamiltonian} .
Furthermore, we can write a first integral of the problem, in terms of the conserved quantities $G$ and $J={\bf G}\cdot {\bf T}$:
\begin{equation}
G^2=\dot f^2_\kappa +(f_\kappa \kappa -f-\Phi)^2+\frac{J^2}{f_\kappa^2}.
\end{equation}
The presence of the long range force, into the effective potential $V(\kappa, \Phi)= (f_\kappa \kappa -f-\Phi)^2$,  does not allow their interpretation
as a central potential $V(\kappa)$ as in the elastic case.

\subsection{The Euler elastica}
In the case of plane curves  with bending energy $f(\kappa)=\kappa^2/2$, the Euler-Lagrange equation reduces to:
\begin{equation}
\ddot\kappa +\kappa\left( \frac{\kappa^2}{2} - \Phi\right)=-{\bf n}\cdot \nabla\Phi,
\end{equation}
this was found in a dynamical context in ref.  \refcite{gold}.
The corresponding first order integral equation can be written as
\begin{equation}
\dot\kappa^2+\left( \frac{\kappa^2}{2}-\Phi \right)^2= G^2,
\end{equation}
and for the particular case of electrostatic interaction
it is given  by
\begin{equation}
\dot\kappa^2+\left( \frac{\kappa^2}{2}-\int  \frac{ds'}{|{\bf X}(s)-{\bf X}(s')|  } \right)^2= G^2,
\end{equation}
in terms of the embedding function ${\bf X}(s)$, it is an integro-differential equation.

\section{Surfaces}\label{Surfaces}

\subsection{Shape equation}

The geometric description of the surface is constructed considering that it
is  embedded in ${\bf R}^3$. We parameterized it by local coordinates $\xi^a$, $a=1, 2$ , through ${\bf x}={\bf X}(\xi^a)$.  Now
there are two tangent vectors ${\bf e}_a=\partial_a {\bf X}$ to the surface, and  the  unit normal vector field ${\bf n}$ is defined by ${\bf e}_a\cdot {\bf n}=0$
and  ${\bf n}\cdot {\bf n}=1$. The induced metric on the surface is defined by  the symmetric tensor
$g_{ab}={\bf e}_a\cdot {\bf e}_b$, whose inverse is denoted $g^{ab}$. We denote $\nabla_a$ the covariant derivative compatible with the induced metric.

The model of the surface that we consider is the sum of two terms, a natural generalization of equation (\ref{modt}) for curves
\begin{equation}
{\mathscr H}=\int dA\, f(K) +\frac{1}{2} \iint dA\, dA' \, h(r)\label{modsur},
\end{equation}
where $dA=\sqrt{g}\,\, d^2\xi$ is the infinitesimal area element on the surface, $g$ being the determinant of the induced metric.

The first term in equation (\ref{modsur}), $\mathscr{H}_1$, is the elastic energy of the surface;  $f(K)$ is a scalar function constructed with the geometry of the surface.  $K=g^{ab}K_{ab}$ the mean curvature in terms of the second fundamental form $K_{ab}=-\nabla_a {\bf e}_b\cdot {\bf n}$.  The second term in equation (\ref{modsur}), $\mathscr{H}_2$,
involves the external potential $h(r)$, similar to the case of curves, section (\ref{Curves}).  The functional $\mathscr{H}$,  is invariant under reparameterizations and euclidean motions
of the surface.

Given the invariance under reparameterizations of the energy, only normal deformations $\delta{\bf X}\cdot {\bf n}$ play a role in the Euler-Lagrange
equations \cite{jemstress}.
Deforming the first term of the energy (\ref{modsur}), we can obtain \cite{jemstress}
\begin{equation}
\delta {\mathscr H}_1= \int dA\, {\cal E}\, \delta{\bf X}\cdot {\bf n}+\int dA\, \nabla_a Q^a\label{DF1},
\end{equation}
where $\cal E$ is the Euler-Lagrange derivative of the elastic model and $Q^a$ the corresponding Noether charges.
The remaining deformation of the energy $\mathscr H$ in eq.(\ref{modsur}) involves the external potential.  We obtain
\begin{equation}
\delta {\mathscr H}_2= \int\ dA\,\left(  K \Phi\, +{\bf n}\cdot \nabla\Phi\right)   \delta{\bf X}\cdot {\bf n}+\int dA\, \nabla_a \left( g^{ab}\delta {\bf X}\cdot {\bf e}_b\Phi \right),\label{DF2}
\end{equation}
where we have defined $\Phi=\int dA'\, h(r)$.
The shape equation can be found by setting $\delta {\mathscr H}_1 +\delta{\mathscr H}_2=0$. From equations (\ref{DF1}) and (\ref{DF2}) we have that the Euler-Lagrange derivative is
\begin{equation}
{\mathscr E}={\cal E}+K\Phi+{\bf n}\cdot\nabla\Phi,
\end{equation}
in equilibrium, the elastic forces are balanced with the external long range force, ${\cal E}=-K\Phi -{\bf n}\cdot\nabla\Phi$. In the case of minimal surfaces where $K=0$, there is no contribution of the second term to the shape equation.

\subsection{Translational and rotational invariance}

We can write the total deformation of the energy in the form
\begin{equation}
\delta {\mathscr H}=\int dA \left( {\cal E}+K\Phi+{\bf n}\cdot\nabla\Phi\right)\delta{\bf X}\cdot {\bf n} +
\int dA\, \nabla_a \left( Q^a+ g^{ab}\delta {\bf X}\cdot {\bf e}_b\Phi \right). \label{ddtt}
\end{equation}
By doing an infinitesimal translation $\delta{\bf X}={\bf a}$ we obtain that
\begin{equation}
\delta{\mathscr H}={\bf a}\cdot \int dA\, \left[ \left(  {\cal E}+K\Phi+{\bf n}\cdot\nabla\Phi\right) {\bf n}
- \nabla_a \left( {\bf f}^a-{\bf e}^a\,\Phi  \right)\right],
\end{equation}
where the relation between the Noether charge and the elastic forces, $Q^a=-{\bf a}\cdot {\bf f}^a$, was used.
Thus, on a surface in equilibrium,  ${\bf h}^a= {\bf f}^a-{\bf e}^a\, \Phi$ is a conserved vector field, as a consequence of invariance under translations.
Using the decomposition in the local basis,  ${\bf f}^a=f^{ab}{\bf e}_b+f^a{\bf n}$, we can write
\begin{eqnarray}
\nabla_a f^{ab}+K^b{}_a f^a&=& \nabla^b\Phi,\nonumber\\
\nabla_a f^a-K_{ab} f^{ab}&=&{\cal E}+{\bf n}\cdot \nabla\Phi,
\end{eqnarray}
comparing with eq.(\ref{ge}) we can see that they match for the case of curves.

Let us see the consequences of the invariance of the energy under rotations.  Let  $\delta{\bf X}=\boldsymbol{\Omega} \times {\bf X}$ be an infinitesimal rotation
in the deformation of the energy equation (\ref{ddtt}). We have
\begin{equation}
\delta{\mathscr H}=\boldsymbol{\Omega}\cdot \int dA \left[ {\mathscr E}\, {\bf X}\times {\bf n}-\nabla_a {\bf M}^a\right],
\end{equation}
the invariance under rotations of the energy implies that ${\mathscr E}\, {\bf X}\times {\bf n}=\nabla_a {\bf M}^a$. Thus, in equilibrium ${\bf M}^a$ is a conserved
vector field on the surface  and it is identified with the torque. We can decompose this vector  in an orbital  torque  plus a local one, in the form
${\bf M}^a={\bf X}\times \left( {\bf f}^a -{\bf e}^a \Phi \right)+{\bf s}^a$  where the local torque satisfies $\nabla_a {\bf s}^a={\bf f}^a \times {\bf e}_a$.

\subsection{Lipid membranes}

In this case\cite{helfrich} the energy is quadratic in the mean curvature $f(K)=\frac{K^2}{2}$. The Euler-Lagrange derivative and the conserved Noether charge
of the elastic energy are modified as follows
\begin{eqnarray}
{\mathscr E}&=&-\nabla^2 K+ K\left( 2K_{G}-\frac{1}{2}K^2\right) +K\Phi + {\bf n}\cdot\nabla\Phi \nonumber\\
{\bf h}^a&=& K\left( 2K^{ab}-Kg^{ab}\right){\bf e}_b -2\nabla^a K {\bf n} +\Phi\,  {\bf e}^a
\end{eqnarray}
where $2K_{G}={\cal R}$, is the  curvature scalar of the surface.

\section{\bf Constraining curves on surfaces. Auxiliary variables.}\label{Auxiliary}

It is possible to impose the relations (\ref{sf}) directly in the energy (\ref{modt}) by considering them as constraints of the variational problem, and then fixing them through the introduction of indeterminate Lagrange multipliers \cite{Aux}. In this way,  the curvature $\kappa$ and $\mathbf{X}$ can be varied independently.  In order to confine the curve into a surface ${\bf x}={\bf Y}(\xi^a)$ we have to add in equation
(\ref{modt}) this constraint in the form
\begin{equation}
{\cal H}_c={\cal H}+\int ds\, \boldsymbol{\lambda}\cdot \left[ {\bf X}(s)-{\bf Y}(\xi^a(s))\right].
\end{equation}
As in ref. \refcite{pablo1} we take the variation with respect to ${\bf X}$  obtaining the following relation
\begin{equation}
\delta_{\bf X}{\cal H}_c =\int ds\, \left( \dot {\bf F} +\boldsymbol{\lambda}+\nabla\Phi \right)\cdot \delta{\bf X}\label{ddi},
\end{equation}
where $\nabla\Phi$ denotes the euclidean gradient,  and the multiplier ${\bf F}$ is given by
\begin{equation}\label{ef}
\mathbf{F} = \left( -\lambda_T-\kappa f_{\kappa}\right) \mathbf{t} + \left( \frac{\tau}{\kappa}\dot{f}_{\tau} + \dot{f}_{\kappa} \right)\mathbf{n}_1 +
\left( \tau f_{\kappa} - \kappa f_{\tau} - \frac{d}{ds}\left( \frac{\dot{f}_{\tau}}{\kappa}\right) \right)\mathbf{n}_2,
\end{equation}
$\lambda_T$ being the Lagrange  multiplier that enforces the curve to be parameterized by its arclength.
From equation (\ref{ddi}), we see that along a curve in equilibrium, $\dot {\bf F} +\nabla\Phi=-\boldsymbol\lambda$. We can also see that $\boldsymbol\lambda$ is orthogonal
to the surface and therefore we can write,  $\boldsymbol\lambda=-\lambda{\bf n}.$ As a consequence the tangential projection vanishes identically, i.e.
$\left( \dot{\bf F} +\nabla\Phi\right)  \cdot {\bf e}_a=0$, in particular, projection along the tangent $\bf t$ determines the multiplier $\lambda_T$  in equation (\ref{ef}) to be:
\begin{equation}
\lambda_T=f+\Phi -2\kappa f_{\kappa}-\tau f_{\tau} +\sigma,
\end{equation}
where $\sigma$ is a constant  that tell us that  the length of the curve is fixed.  Thus,  the external potential $\Phi$ contributes to the effective tension.
Projections of $\dot{\bf F} +\nabla\Phi$, along the normal vectors ${\bf n}_i$ give the Euler-Lagrange equations. We have
\begin{eqnarray}
(\dot{\bf F} +\nabla\Phi )\cdot {\bf n}_1&=& \ddot f_\kappa +(\kappa^2-\tau^2) f_{\kappa}+2\tau\frac{d}{ds}\left( \frac{\dot f_\tau}{\kappa}\right) +\dot\tau\left( \frac{\dot f_\tau}{\kappa} \right)
+2\kappa\tau f_\tau \nonumber\\
 &-&\kappa \left( f+\Phi +\sigma\right) +{\bf n}_1\cdot \nabla\Phi,  \\
(\dot{\bf F}+\nabla\Phi )\cdot {\bf n}_2&=& 2 \frac{d}{ds}\left( \tau f_\kappa \right)-\dot\tau f_\kappa -\frac{d^2}{ds^2}\left( \frac{\dot f_{\tau}}{\kappa}\right)- \frac{d}{ds}(\kappa f_\tau ) \nonumber \\
&+&\frac{\tau^2\dot f_\kappa}{\kappa}  + {\bf n}_2\cdot\nabla\Phi,
\end{eqnarray}
these equations reproduce the $E_i$ derivatives of equations (\ref{euler}) in the case of curves without restriction.  Then we can write $(\dot{\bf F} +\nabla\Phi )=E_i{\bf n}_i$.
If we identify the normal to the surface to be  the principal normal  of the curve, ${\bf n}={\bf n}_1$,  then we can write
\begin{equation}
\dot{\bf F}+\nabla\Phi=E_{\bf n} {\bf n}+E_{\bf l}{\bf l}, \label{deco}
\end{equation}
where the unit vector field ${\bf l}={\bf n}_2$ is defined by the conditions ${\bf l}\cdot {\bf t}=0$ and ${\bf l}\cdot {\bf n}=0$.  From equation (\ref{deco}), the Euler Lagrange equation,
 $E_{\bf l}=0$ and the normal force, $-\lambda$, can be recognized as

\begin{eqnarray}
E_{\bf l}&=& (\dot{\bf F} +\nabla\Phi)\cdot{\bf l},\\
-\lambda &=&(\dot{\bf F}+\nabla\Phi)\cdot {\bf n}.
\end{eqnarray}
That is, the long range force contributes to the normal geometric forces adding the terms $ \kappa\Phi - {\bf n}\cdot\nabla\Phi $.

With this formulation one can also consider the long range interaction between points on the surface with points on the curve, this can be achieved by adding the term
\begin{equation}\label{Surf-Curv}
 \iint \, ds\, dA'\,  h(R),
\end{equation}
where now $R=|{\bf X}(s)-{\bf Y}\left(\xi^a(s')\right)|$. This term considers the contributions to the interaction of the surface on the curve, for every point on the curve. Notice that the back-reaction that the curve has on the surface is not considered. The force satisfies
\begin{equation}\label{el}
\dot{\mathbf{F}} + \nabla\left(  \Phi  + \tilde{\Phi}\right)  =-\boldsymbol{\Lambda}(s) ,
\end{equation}
where $\tilde{\Phi}=\int dA'\, h(R)$ is the contribution of the potential on the surface.
Bearing in mind that now we can vary with respect to the coordinates on the surface $\xi^a$, that leads us to
\begin{equation}\label{varxi}
 \delta_{\xi} {\cal H} = -\int ds\, \boldsymbol{\Lambda}(s)\cdot\mathbf{e}_a\delta\xi^a - \int ds\, \nabla\tilde{\Phi}\cdot \mathbf{e}_a\delta \xi^a,
\end{equation}
such that $(\boldsymbol{\Lambda}(s) + \nabla\tilde{\Phi})\cdot\mathbf{e}_a = 0$. This means that the only force that appears on the surface is $\nabla\Phi$.

From the variation of ${\cal H}$ with respect to the remaining variables one can obtain the Lagrange multipliers:
\begin{equation}\label{Lmulti}
    \Lambda_{\kappa} = {\cal H}_{\kappa},\qquad \Lambda_{\tau} = {\cal H}_{\tau}.
\end{equation}
Using the properties of stationarity and Frenet-Serret equations, the following relation for the components of the multiplier $\mathbf{F}$, which corresponds to the force, can be derived:
\begin{equation}\label{efe}
   \mathbf{F} = \left( -\lambda - \kappa{\cal H}_{\kappa}\right)\mathbf{t} + \left( \frac{\tau}{\kappa}\dot{{\cal H}}_{\tau} + \dot{{\cal H}}_{\kappa} \right)\mathbf{n}_1 + \left( \tau{\cal H}_{\kappa} - \kappa {\cal H}_{\tau} - \frac{d}{ds}\left(\kappa^{-1}\dot{{\cal H}}_{\tau}\right)  \right)\mathbf{n}_2,
\end{equation}
where only the multiplier $\lambda$ is missing. To find  $\lambda$ one can take the derivative of (\ref{efe}), and compare its components with those obtained in (\ref{el}), which are clearly different form zero. With this, one can find the following relations:
\begin{eqnarray}
  \dot{\mathbf{F}}\cdot \mathbf{t} &=& -\left(\kappa{\cal H}_{\kappa}\right)\dot{} - \tau\dot{{\cal H}}_{\tau}- \kappa\dot{{\cal H}}_{\kappa} - \dot{\lambda} = -\dot{\Phi}, \label{ft}\\
  \dot{\mathbf{F}}\cdot \mathbf{n}_1 &=& \ddot{{\cal H}_{\kappa}} - \left(\kappa^2 + \tau^2\right){\cal H}_{\kappa} +     2\tau   \left(\kappa^{-1}\dot{{\cal H}}_{\tau}\right)\dot{} + \dot{\tau} \left(\kappa^{-1}\dot{{\cal H}}_{\tau}\right) + \kappa\tau{\cal H}_{\tau} -\kappa\lambda \nonumber  \\
  &=& - \left(\nabla\Phi + \boldsymbol{\Lambda} + \nabla\tilde{\Phi}\right)\cdot \mathbf{n}_1, \label{fn1}\\
  \dot{\mathbf{F}}\cdot \mathbf{n}_2 &=& \left( \tau{\cal H}_{\kappa} \right)\dot{} - \left( \kappa^{-1}\dot{{\cal H}}_{\tau} \right)\ddot{} - \left(\kappa{\cal H}_{\tau} \right)\dot{} + \tau\left(\frac{\tau}{\kappa}\dot{{\cal H}}_{\tau}+\dot{{\cal H}}_{\kappa}\right)\nonumber\\
  &=& -\left(\nabla\Phi + \boldsymbol{\Lambda} + \nabla\tilde{\Phi}\right)\cdot \mathbf{n}_2. \label{fn2}
\end{eqnarray}
From (\ref{ft}) one can integrate to obtain $\lambda$, noticing that this expression is a total derivative on $s$:
\begin{equation}\label{lambda}
    \lambda = {\cal H} - 2\kappa{\cal H}_{\kappa} - \tau{\cal H}_{\tau} + \Phi + \sigma,
\end{equation}
where $\sigma$ again fixes the length of the curve. With (\ref{lambda}), the multiplier $\mathbf{F}$ is determined as:
\begin{eqnarray}
\mathbf{F} &=& \left( \kappa{\cal H}_{\kappa} + \tau{\cal H}_{\tau} - {\cal H} - \Phi - \sigma \right)\mathbf{t} + \left( \frac{\tau}{\kappa}\dot{{\cal H}}_{\tau} +
\dot{{\cal H}}_{\kappa} \right)\mathbf{n}_1\nonumber\\
&+& \left( \tau{\cal H}_{\kappa} - \kappa {\cal H}_{\tau} - \left(\kappa^{-1}\dot{{\cal H}}_{\tau}\right)\dot{}  \right)\mathbf{n}_2.\label{f}
\end{eqnarray}
Amusingly, both the force that keeps the curve on the surface and the interaction with the membrane, do not appear in this expression; only the potential due to the interaction of the curve itself appears in the tangential component.

\section{Concluding Remarks}
\label{Remarks}

The interest on studying charged polymers and surfaces comes because in nature, there are some systems, such as the polyelectrolytes \cite{Poly}, that can be modeled as a charged chains which require a description involving long-range electrostatic forces.

In this work we have studied curves and surfaces under the influence of long-range and elastic forces, both from  the classical differential geometry perspective.
Besides the usual elastic terms, we introduce the long-range interactions as an integral of the corresponding potential, which takes into account the interaction between all points of the curve (\ref{modt}).
The corresponding shape equations were obtained form a variational principle, considering small deformations.

In the case of curves the resulting Euler-Lagrange equations can be written as the sum of the usual elastic term and normal projections of the external force. Amusingly, in one normal directions there are a contribution of the bare potential, which further is coupled with the curvature. We can interpret this term in (\ref{euler}) as a minimal coupling. Given the translational and rotational invariance of the model, we were able to write the equations for the force and torque, noticing that the long-range potential $\Phi$ contributes directly to the orbital torque, unlike the force and the local torque, where it only contributes through its gradient. In the case where the elastic force is quadratic on the curvature, the well known Euler \textit{elastica} model, the corresponding equation for $\kappa$ becomes an integro-differential equation that will be solved numerically elsewhere.
For the charged surfaces we perform the same analysis straightforwardly, and found a similar minimal coupling, now with the mean curvature, and the projection of the long-range force in the normal direction; both in the Euler-Lagrange derivative.

We also consider the case of a curve constrained to lie on a surface, this analysis was done by introducing Lagrange multipliers for the constraints of the system, recovering the previous results, and obtaining additional information for the normal force on the curve. Moreover, with this formalism, we could write the corresponding interaction in the case where both the surface and the curve were charged.

To integrate all the resulting equations, numerical methods must be used, although it was not the main task of this work, we are in the process of implementing this in a subsequent study.

\section*{Acknowledgments}

We would like to thank Jemal Guven  for many useful discussions on the topics here addressed. This work was partially supported (GCA) by project PROMEP 47510283.

\end{document}